\documentclass[final,1p,times]{elsarticle}
\usepackage{graphicx}% Include figure files
\usepackage{dcolumn}% Align table columns on decimal point
\usepackage{bm}% bold math
\usepackage[english]{babel}
\usepackage{t1enc}
\usepackage{subfigure}
\usepackage{times}

\journal{Astroparticle Physics}

\begin{document}
\begin{frontmatter}
\title{Source position reconstruction and constraints on the galactic magnetic field from ultra-high energy cosmic rays}
\author{Geraldina Golup, Diego Harari, Silvia Mollerach and Esteban Roulet}
\address{CONICET and Centro At\'omico Bariloche \\ Av. Bustillo 9500, 8400, S. C. de Bariloche, Argentina}
\begin{abstract}
We study the possibility to reconstruct the position of ultra-high energy cosmic ray sources and some
properties of the magnetic field along the line of sight towards them in the case that several events from the
same source are detected. By considering an illustrative model for the galactic magnetic field, including both a
regular and a turbulent component, we estimate the accuracy that can be achieved in the reconstruction. We
analyse the effect of the experimental energy and angular resolutions on these results. We show that if about
ten events with energies above 30 EeV are detected coming from the same source, it should be possible to
reconstruct the source position with an accuracy of 0.5$^{\circ}$ and the integral of the orthogonal component
of the magnetic field along the line of sight with an accuracy of 0.6 $\mu$G kpc Z$^{-1}$ (with Z the electric
charge of the particles).
\end{abstract}
\begin{keyword}
High Energy Cosmic Rays
\PACS 98.70.Sa
\end{keyword}

\end{frontmatter}
%%%%%%%%%%%%%%%%%%%%%%%%%%%%%%%%%%%%%%%%%%%%%%%%%%%%%%%%%%%%%%%%%%%%%%%%%%%%%%%%%%%%%%%%%%%%%%
\section{Introduction}
The recent discovery by the Pierre Auger Collaboration \cite{science} that the arrival directions of the
ultra-high energy cosmic rays are correlated with the nearby extragalactic matter distribution marks a first
step in the cosmic ray astronomy era. Out of their 27 highest energy events, 20 correlate with the position of
nearby active galactic nuclei (AGN), with two events near Cen A, our closest AGN. Previously, the combined data of AGASA, Haverah Park, Volcano Ranch and Yakutsk showed signals of clustering, including eight doublets and two triplets within 4$^{\circ}$ angular separation \cite{uchihori}. It is to be expected that with increased statistics future data will contain evidence of multiplets, i.e. cosmic rays of different energies coming from the same point-like source.  From the arrival direction and energy of multiplets one can extract not only the position of the source but also valuable information about the magnetic field along the line of sight towards it. In this work we analyse how to reconstruct this information considering the effect that energy and angular experimental resolutions have on the results and estimate the accuracy that can be achieved.\\

Despite the considerable observational efforts done, the magnetic field of the Galaxy and the extragalactic
ones are still poorly known. This, together with the fact that the cosmic ray composition at ultra-high energies
is also poorly known, makes it difficult to identify the cosmic ray sources, as it is not possible to predict
the deflection of the cosmic ray trajectories in their way from the sources to the Earth. Propagation of charged particles through the galactic magnetic field has been studied in detail previously in many references. These studies are usually done backtracking antiparticles leaving the Earth with an initial direction pointing to the arrival direction of the cosmic ray particle \cite{stanev,tanco,toes,tinyakov,prouza,takami}. Alternatively, the forward tracking technique, consisting of tracking the trajectories of a large number of particles leaving the source and keeping the arrival directions of those particles that arrive to a given neighborhood of the Earth, has been used in \cite{alvarez}. Bisymmetric \cite{stanev,toes,tinyakov,prouza,takami,alvarez} and axisymmetric \cite{stanev,tanco,toes,takami} magnetic field models have been considered, as well as a dipole \cite{prouza,takami,alvarez} and toroidal components  \cite{prouza}. A turbulent magnetic field component was included in \cite{prouza,alvarez}. These studies show that
the magnitude and direction of the resulting deflections are strongly dependent on the details of the magnetic field model considered.\\

The observation of several events from the same source would open the possibility to actually measure the integral of the component of the magnetic field orthogonal to the cosmic ray trajectory, providing a new constraint to galactic magnetic field models, besides the possibility to locate more accurately the source position. Magnetic fields not only
deflect the trajectories of charged particles but they can also amplify or demagnify the flux arriving from the
source, modifying its spectrum. A detailed analysis of magnetic lensing effects can be found in refs. \cite{toes,sign,turb,spec}.\\

In this paper we will use simulations of sets of events arriving to the Earth from randomly located sources in
the sky after travelling through the galactic magnetic field, which we model using a regular and a turbulent
component that aim to reproduce the general characteristics of the observational results. For each hypothetical source direction in the sky we consider several cosmic ray particles with different energies for which we
compute the arrival directions to the Earth. The energies of the simulated events are generated using a power
law at the source, which is then modified according to the lensing effects expected in each sky direction
\cite{toes}. For each simulated set of events from a source we then reconstruct the original direction of the
source and the integral of the orthogonal component of the magnetic field, comparing the results with the actual
values to estimate the uncertainty of the method and the effect of the experimental error in the
determination of the energy and the arrival direction.\\

The paper is structured as follows: in section II we describe the general known characteristics of the galactic magnetic field and
review some magnetic lensing facts. Then, in section III, we explain the method used for the analysis and
reconstruction and in section IV we present and discuss the accuracy that can be achieved in the reconstruction
of the magnetic field integral and of the source position for realistic situations. Finally, section V is a
summary of the results and contains our conclusions.\\

%%%%%%%%%%%%%%%%%%%%%%%%%%%%%%%%%%%%%%%%%%%%%%%%%%%%%%%%%%%%%%%%%%%%%%%%%%%%%%%%%%%%%%%%%%%%%
\section{Galactic Magnetic Field and Magnetic Lensing}
The present knowledge of the galactic magnetic field is far from complete. There are several observational
methods for determining it: Zeeman splitting, polarized thermal emission from dust in clouds, polarization of
starlight, synchrotron radio emission and Faraday rotation of polarized sources \cite{han1,beck}. When using
rotation measures of galactic pulsars to constrain the three most widely used theoretical models for the magnetic
field in the Galactic disk, namely the circular, the axisymmetric and the bisymmetric field models, none of
them can consistently fit all the data, although the bisymmetric model appears to perform better than the
others \cite{han2}. The observation of ultra-high energy cosmic ray multiplets could provide a new handle to
probe the galactic magnetic field.\\

The galactic magnetic field has a large scale regular component and a turbulent component present on galactic
scales. From the analysis of the polarization of starlight it is known that the local regular component is
approximately parallel to the galactic plane and follows the local spiral arms \cite{heiles}.
The regular component has a local value of $B_{reg}\simeq 2\ \mu$G according to Faraday rotation measures
\cite{han3} (although this local value could be underestimated if the fluctuations in magnetic field and electron density were anticorrelated \cite{beck2}). The local value of the regular field derived from synchrotron measurements is larger, $B_{reg}\simeq 4\ \mu$G \cite{beck}, but it could be overestimated due to the anisotropy of the turbulent magnetic field \cite{beck2,cowsik,sarkar}. According to some models the regular magnetic field has reversals in direction between neighboring arms. From Faraday rotation data there is evidence that in the inner arms magnetic fields are counterclockwise when viewed from the North Galactic pole, while in the local region the fields in the interarm regions are clockwise \cite{han1}. Rotation measures also indicate that the magnetic field in the galactic halo appears to be antisymmetric with respect to the galactic plane and such a field could be produced by an A0 dynamo mode \cite{han1}. However, it is argued that the superposition of a disk field with even parity and a halo field with odd parity cannot be explained by classical dynamo theory \cite{beck}.\\

The random component has a root mean square amplitude of $B_{rms} \simeq (1-2) B_{reg}$ and a typical coherence length $L_c\simeq$100 pc \cite{rand, ohno}. The regular component, being coherent on scales much larger than $L_c$, produces the dominant effect on the deflection of high energy charged particles travelling through the Galaxy. In addition, the presence of extragalactic magnetic fields could also be relevant for the propagation of cosmic rays. However, their strength is very uncertain (and likely subdominant as regards the deflecting power) and we will not consider their effects here.\\

\begin{figure}[!htb]
\begin{center}
\subfigure[\label{fig1:a}]{\includegraphics[scale=0.65]{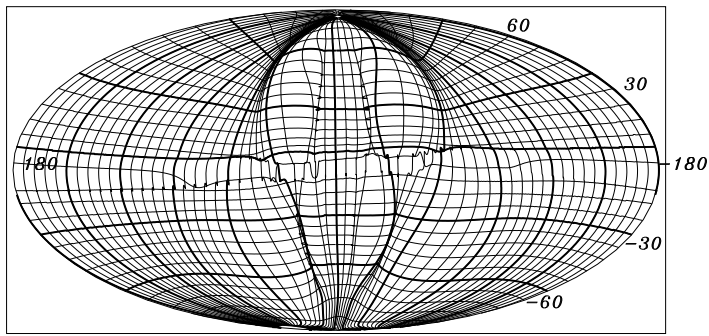}}
\subfigure[\label{fig1:b}]{\includegraphics[scale=0.65]{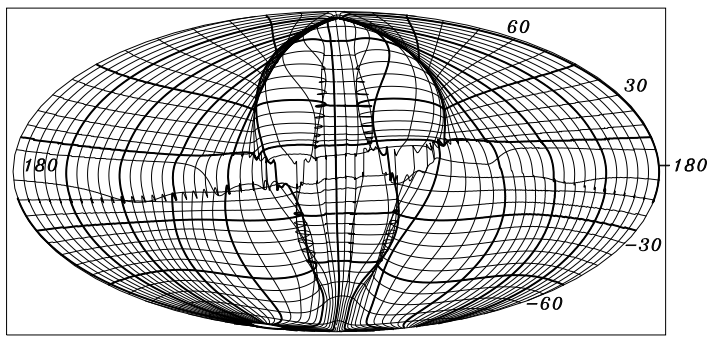}} \caption[]{`Sky sheets': directions of
incoming cosmic rays in the halo that correspond to a regular grid of arrival directions at Earth, adopting the BSS-S
magnetic field configuration, for particles with $E/Z$ = 30 EeV (a) and 20 EeV (b). An Aitoff projection in galactic
coordinates centered in the Galactic center is used.}
\end{center}
\end{figure}

In this paper we consider an illustrative galactic field model that includes some of the observed features as field reversals and reproduce the local field strength. The regular component of the galactic magnetic field is modeled with a bisymmetric field with even symmetry (BSS-S) with structure and strength very similar to those used in \cite{stanev} but smoothed out as described in \cite{toes}. In this model the galactic magnetic field reverses its sign between the arms of the Galaxy and the field is symmetric with respect to the Galaxy's mid-plane. The local value of the field is taken as $2\ \mu$G. For the dependence on $z$ a contribution coming from the galactic disk and another one from the halo are considered:
\begin{equation}
\vec B_{reg}(x,y,z)=\vec B_{reg}(x,y,z=0) \left(\frac{1}{2\cosh(z/z_1)}+\frac{1}{2\cosh(z/z_2)}\right)
\end{equation}
with $z_1$= 0.3 kpc and $z_2$= 4 kpc. These values of the parameters $z_1$ and $z_2$
lead to a similar dependence with the height above the galactic plane $z$ as that obtained using
an exponential profile $e^{-|z|/z_{0}}$ with scale heights $z_0$= 1.2 kpc for $|z|< 0.5$ kpc and $z_0$= 4 kpc
for $|z|> 0.5$ kpc, as considered in \cite{stanev}, but the expression adopted in eq. 1 allows to avoid
the presence of singular derivatives on the plane and when matching the disk and halo profiles.
To model the turbulent component of the magnetic field
we use a Gaussian random field with zero mean and root mean square value $B_{rms}$ equal to $2\ \mu$G.
The coherence length $L_c$ adopted is 100 pc. For the dependence on $z$ of the turbulent component we consider
a contribution coming from the disk and another one from the halo with the same scales as in the regular component.\\

The deflections caused by the magnetic field can lead to lensing phenomena, such as the energy dependent
magnification and demagnification of the flux that modifies the energy spectrum of the source, or to the
appearance of secondary images \cite{toes,sign,turb}. The magnitude of the flux amplification depends on the
arrival direction, on the ratio between the energy and charge $E/Z$ of the cosmic ray and on the magnetic field
model. The formation of multiple images and the flux magnification can be understood pictorially plotting for a
regular grid of arrival directions at Earth the corresponding directions from which the particles arrived to the
galactic halo. This is shown in Figure 1 for the model of the regular magnetic field configuration that is considered in this paper and for particles with $E/Z$ = 30 EeV (top panel) and 20 EeV (bottom panel), where 1
EeV $\equiv 10^{18}$ eV. One may picture this distorted image of the sky seen from the Earth as a sheet (the
`sky sheet') that can be stretched and folded. A source located in a fold of this sky sheet will have multiple
images, i.e. cosmic rays of the same energy can arrive to the Earth from several different directions. Moreover,
the flux coming from a source in a region where the sheet is stretched will appear demagnified while that from a
source in a compressed region will appear magnified. As the sky sheet changes with the energy, the spectrum observed from a given source is different from the emitted one \cite{toes}.\\

Magnetic lensing phenomena also appear for turbulent fields \cite{turb}. In this case, multiple images appear
below a critical energy $E_c$ such that typical transverse displacements among different paths after travelling
a distance $L$ in the turbulent field become of the order of the correlation length of the random magnetic field
($\delta_{rms} \sim L_c/L$). This critical energy is typically higher than the corresponding one for the regular
magnetic field, but the folds produced by the random field near the critical energy are on a much smaller
angular scale. For decreasing energies, the fraction of the sky covered with folds increases, however the
magnification peaks become increasingly narrower and for $E<E_{c}/3$ their integrated effect becomes less
noticeable.\\

%%%%%%%%%%%%%%%%%%%%%%%%%%%%%%%%%%%%%%%%%%%%%%%%%%%%%%%%%%%%%%%%%%%%%%%%%%%%%%%%%%%%%%%%%%%%%
\section{Reconstruction of the source position and the magnetic field}
Charged particles of different energies coming from the same source suffer different deflections in their way
through the Galaxy and are thus observed with different arrival directions. If deflections are small, the
arrival direction $\vec \theta$ of a particle with energy $E$ is related to the source direction $\vec \beta$ by
\begin{equation}\label{small}
\vec{\beta}=\vec{\theta} + \frac{\vec{F}(\vec{\theta})}{E},
\end{equation}
where $\vec{F}$ is the integral along the line of sight of the perpendicular component of the magnetic field
$\vec B$ times the charge $Ze$ of the particle
\begin{equation}
\vec{F}(\vec{\theta})=Ze\int^{L}_{0} {\rm d} \vec{l} \times \vec{B}(\vec{l}).
\end{equation}
To analyse the correlation between arrival direction and energy, we first fit in the tangent plane to the celestial
sphere in the direction of the events (using coordinates $\alpha_{1}$ and $\alpha_{2}$) a straight line,
$\alpha_{1}=a_1+a_2 \alpha_{2}$, to the event coordinates in order to determine the direction of the deflection.
This direction should coincide with the direction of $\vec{F}$. Then, we rotate to new coordinates ($\theta_1,\theta_2$) along and orthogonal to the deflection direction. The procedure is illustrated in Figure 2.\\

\begin{figure}[!h]
\begin{center}
\includegraphics[scale = 0.85]{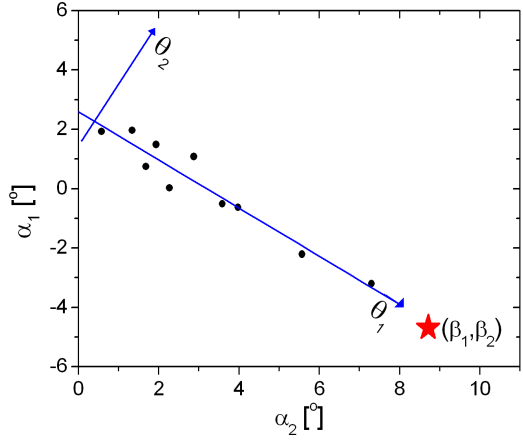}
\caption[]{An example of the rotations performed. In the tangent plane to the celestial
sphere in the direction of the events, coordinates ($\alpha_{1}$,$\alpha_{2}$), we perform a linear fit to the event coordinates in order to determine the direction of the deflection. The new coordinates are named ($\theta_1,\theta_2$), along and orthogonal to the deflection direction. In these new coordinates, the source position es ($\beta_1,\beta_2$).}\label{figura 0}
\end{center}
\end{figure}

If the deflections are small we can approximate $\vec{F}(\vec{\theta})$ as a constant value, lets say the value in
the source direction $\vec{F}(\vec{\beta})$, and fitting a linear relation between $\theta_1$ and $1/E$,
\begin{equation}
\theta_1 = \beta_1 - \frac{F_{\theta_1}(\vec \beta)}{E},
\end{equation}
we can obtain the component $\theta_1$ of the position of the source, $\beta_1$, and the component in the
direction of the coordinate $\theta_1$ of the integral $\vec F$ evaluated at the position of the source,
$F_{\theta_1}(\vec \beta)$. The position of the source in these coordinates will be $\vec{\beta} =(\beta_1,0)$
as the deflection is essentially along the coordinate $\theta_1$.\\

The fact that particles of different energies follow different paths along the galactic magnetic field has as a
consequence that $\vec{F}(\vec{\theta})$ is not actually constant, so departures from the linear relation are
expected, specially for the lower energy events, which follow more bent paths. In these cases the next term in
the expansion can be relevant
\begin{equation}
\beta_1 = \theta_1 + \frac{1}{E} \left[F_{\theta_1}(\vec{\beta}) + \frac{\partial F_{\theta_1} }{\partial
\theta_1}\bigg{\arrowvert}_{\vec{\beta}} (\theta_1 - \beta_1)\right].
\end{equation}
Assuming that $\left(1 + \frac{1}{E} \frac{\partial F_{\theta_1}}{\partial
\theta_1}\bigg{\arrowvert}_{\vec{\beta}}\right)^{-1} \simeq \left(1 - \frac{1}{E} \frac{\partial F_{\theta_1}
}{\partial \theta_1}\bigg{\arrowvert}_{\vec{\beta}}\right)$, one obtains
\begin{equation}
\theta_1 \simeq \beta_1 - \frac{F_{\theta_1}(\vec{\beta})}{E} +\frac{1}{E^{2}}
F_{\theta_1}(\vec{\beta})\frac{\partial F_{\theta_1}}{\partial \theta_1}\bigg{\arrowvert}_{\vec{\beta}}.
\end{equation}

We will test here the accuracy that can be achieved in the reconstruction of the relevant parameters using this
method. We work with simulations of protons coming from extragalactic point-like sources and propagating through
a toy model of the galactic magnetic field that aims to reproduce the general characteristics of the observational results. These simulations are done backtracking antiprotons leaving the Earth up to a distance where the effect of the galactic magnetic field becomes negligible \cite{toes}. The initial direction of the backtracked antiproton (that will be associated to the arrival direction of protons of the same energy) is recursively adjusted till the final direction points to the source with accuracy better than $10^{-4}$ degrees when only the regular component of the galactic magnetic field is considered and better than $0.3^{\circ}$ when both the regular and turbulent components are considered.\\

We will consider a large number of source directions in order to characterize the deflections expected in
different regions of the sky. For each source direction we simulate 10 events with energies randomly chosen
between 30 and 300 EeV following an $E^{-2}$ spectrum at the source and taking into account the magnification of
each source flux as a function of the energy for that direction. We propagate them through the magnetic field,
keeping track of the arrival direction to the Earth. In order to study the effect of the energy and angular
resolution a random deflection of up to $1^\circ$ is added to the position and a random shift in energy of up to
10$\%$ is introduced. The reconstruction accuracy slightly depends on the particular realization of the
experimental uncertainty, we thus present the results for the mean of 100 different realizations. When the
random component of the galactic magnetic field is introduced the results also slightly depend on the particular
realization of the field, we then present the results of 10 different realizations of the random field
\footnotemark[1] \footnotetext[1]{In a realistic observational situation, the events from one source could be
surrounded by background events from other sources, in which case one can select the candidate source events
using their filamentary structure \cite{mst} and the expected high degree of correlation
between their deflections and $1/E$. A particularly useful tool for this purpose is given by the correlation
coefficient $C(\theta_1, 1/E) = Cov(\theta_1, 1/E)/\sqrt{Var(\theta_1) Var(1/E)}$. Requiring this coefficient to be
sufficiently close to unity for the selected events would ensure their high degree of correlation and hence
a suppressed contribution from background events.}.\\

In each case we perform a linear and a quadratic fit to the relation $\theta_1$ vs. $1/E$ for the events and
obtain the values of $F_{\theta_1}(\vec \beta)$ and the position of the source ($\beta_1,\beta_2$). The actual
value of $\vec{F}(\vec \beta)$ is calculated with a numerical integration along a straight path. By comparing
the reconstructed values of $F_{\theta_1}(\vec \beta)$ and $\vec \beta$ with the actual ones we can estimate the
accuracy of the reconstruction. We also analyse the angle between $\vec{F}(\vec \beta)$ and the direction of the
deflection obtained from the fit to the data, $\epsilon$, which measures the accuracy in the reconstruction of
the direction of $\vec F(\vec \beta)$.\\

In Figure 3 we show the source position and the cosmic ray arrival directions for 100 randomly selected
directions in the sky and in Figure 4 we present the distribution of $F_{\theta_1}(\vec \beta)$ for these
positions. As one can see from these figures, the magnitude of the deflection depends on the region of the sky
where the source is located and has a mean value $\left < F_{\theta_1} \right >$= 2$^{o}$ 100 EeV (1$^{\circ}$100 EeV $\approx$ 1.9 $e$ $\mu$G kpc). Regarding the magnification,
at 30 EeV we find that 19 sources have an amplification smaller than 0.5 and 5 sources have an amplification
larger than 1.5. Moreover, 7 sources have a magnification peak with magnification larger than 3 at some energy
above the 30 EeV threshold. In these peaks, the amplification of the source remains larger than 1.5 for a range
of energies of about 10 to 20 EeV.\\

\begin{figure}[!h]
\begin{center}
\includegraphics[scale = 0.85]{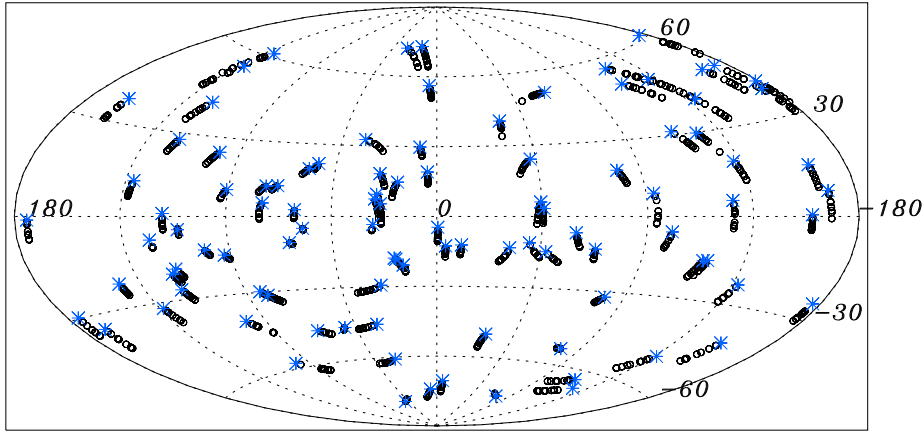}
\caption[]{One hundred sources with randomly selected directions in the sky (asterisks) and ten cosmic rays with
$E/Z\geq 30$ EeV coming from each of these sources (circles) in an Aitoff projection of the celestial sphere in
galactic coordinates. }\label{figura 2}
\end{center}
\end{figure}

\begin{figure}[!htb]
\begin{center}
\includegraphics[scale = 0.85]{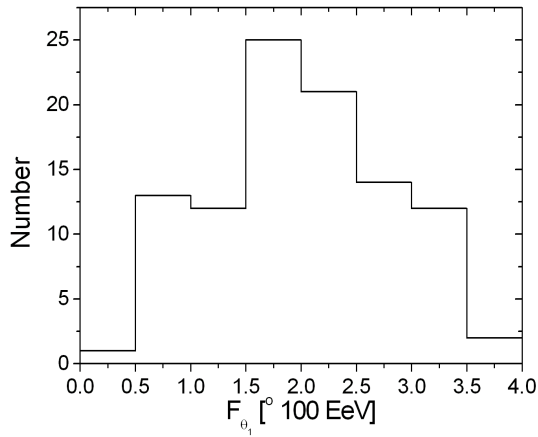}
\caption[]{Distribution of $F_{\theta_1}(\vec \beta)$ obtained with numerical integration for the one hundred sources
considered.}\label{figura 3}
\end{center}
\end{figure}

To illustrate the reconstruction method and its accuracy we first present in detail the results for three
particular source directions which are representative of different types of deflections. The reconstruction
accuracy for the different quantities are summarized in Table 1. The values presented are median values, i.e.
50$\%$ of the realizations of the experimental uncertainties and of the turbulent field have equal or better
accuracy. For each case we first present the results without taking into account any experimental uncertainty.
Thus, these results represent the uncertainties inherent to the method, that are due to the departures of the
actual deflections from the linear or quadratic relations as a function of $1/E$. We then introduce the
experimental uncertainty in the angle and the energy so that it is possible to see its effect on
the reconstruction accuracy.\\

\begin{table}[!hbtp]
\begin{tabular}{lccccc}
\hline \hline
\\
~~~~       &~~~~$\Delta F_{\theta_1}$~~ & ~~ $\epsilon$~~ & ~~ $\Delta \beta_1$~~ & ~~ $\Delta \beta_2$~~ & ~~ $\mid \Delta \vec{\beta} \mid$~~\\
~~~~       &~~~~$[^{\circ} 100\ {\rm EeV}]$~~ & ~~ $[^{\circ}]$~~ & ~~ $[^{\circ}]$~~ & ~~ $[^{\circ}]$~~ & ~~ $[^{\circ}]$~~\\
\\
\hline $Example\ 1:\ (l,b)=(220^{\circ},-60^{\circ})$
\\
\hline
\\
$Regular,\ linear\ fit$          & ~~$0.21$~~  & ~~$1.0$~~  & ~~$0.2$~~  & ~~$0.04$~~ & ~~$0.2$~~ \\
$Regular,\ quadratic\ fit$       & ~~$0.11$~~  & ~~$1.0$~~  & ~~$0.04$~~ & ~~$0.04$~~ & ~~$0.05$~~ \\
$Reg. + error,\ linear\ fit$     & ~~$0.31$~~  & ~~$1.7$~~  & ~~$0.5$~~  & ~~$0.3$~~  & ~~$0.6$~~ \\
$Reg. + error,\ quadratic\ fit$  & ~~$0.75$~~  & ~~$1.7$~~  & ~~$0.6$~~  & ~~$0.3$~~  & ~~$0.7$~~ \\
$Turbulent,\ linear\ fit$        & ~~$0.16$~~  & ~~$1.0$~~  & ~~$0.1$~~  & ~~$0.07$~~ & ~~$0.2$~~ \\
$Turbulent,\ quadratic\ fit$     & ~~$0.22$~~  & ~~$1.0$~~  & ~~$0.1$~~  & ~~$0.07$~~ & ~~$0.2$~~ \\
$Turb. + error,\ linear\ fit$    & ~~$0.28$~~  & ~~$2.3$~~  & ~~$0.4$~~  & ~~$0.3$~~  & ~~$0.5$~~ \\
$Turb. + error,\ quadratic\ fit$ & ~~$0.95$~~  & ~~$2.3$~~  & ~~$0.6$~~  & ~~$0.3$~~  & ~~$0.8$~~ \\
\\
\hline $Example\ 2:\ (l,b)=(55^{\circ},20^{\circ})$
\\
\hline
\\
$Regular,\ linear\ fit$          & ~~$0.28$~~ & ~~$8.6$~~  & ~~$0.2$~~  & ~~$0.1$~~  & ~~$0.3$~~ \\
$Regular,\ quadratic\ fit$       & ~~$0.03$~~ & ~~$8.6$~~  & ~~$0.01$~~ & ~~$0.1$~~  & ~~$0.1$~~ \\
$Reg. + error,\ linear\ fit$     & ~~$0.25$~~ & ~~$21.0$~~ & ~~$0.3$~~  & ~~$0.4$~~  & ~~$0.6$~~ \\
$Reg. + error,\ quadratic\ fit$  & ~~$0.49$~~ & ~~$21.0$~~ & ~~$0.4$~~  & ~~$0.4$~~  & ~~$0.7$~~ \\
$Turbulent,\ linear\ fit$        & ~~$0.27$~~ & ~~$5.2$~~  & ~~$0.3$~~  & ~~$0.1$~~  & ~~$0.3$~~ \\
$Turbulent,\ quadratic\ fit$     & ~~$0.18$~~ & ~~$5.2$~~  & ~~$0.1$~~ & ~~$0.1$~~  & ~~$0.2$~~ \\
$Turb. + error,\ linear\ fit$    & ~~$0.27$~~ & ~~$19.1$~~ & ~~$0.4$~~  & ~~$0.6$~~  & ~~$0.8$~~ \\
$Turb. + error,\ quadratic\ fit$ & ~~$0.64$~~ & ~~$19.1$~~ & ~~$0.5$~~  & ~~$0.6$~~  & ~~$0.9$~~ \\
\\
\hline $Example\ 3:\ (l,b)=(20^{\circ},-40^{\circ})$
\\
\hline
\\
$Regular,\ linear\ fit$          & ~~$0.35$~~ & ~~$10.2$~~ & ~~$0.4$~~  & ~~$0.02$~~ & ~~$0.4$~~ \\
$Regular,\ quadratic\ fit$       & ~~$0.24$~~ & ~~$10.2$~~ & ~~$0.07$~~ & ~~$0.02$~~ & ~~$0.07$~~ \\
$Reg. + error,\ linear\ fit$     & ~~$0.42$~~ & ~~$16.1$~~ & ~~$0.5$~~  & ~~$0.3$~~  & ~~$0.6$~~ \\
$Reg. + error,\ quadratic\ fit$  & ~~$0.42$~~ & ~~$16.1$~~ & ~~$0.4$~~  & ~~$0.3$~~  & ~~$0.6$~~ \\
$Turbulent,\ linear\ fit$        & ~~$0.36$~~ & ~~$7.0$~~  & ~~$0.5$~~  & ~~$0.09$~~ & ~~$0.5$~~ \\
$Turbulent,\ quadratic\ fit$     & ~~$0.28$~~ & ~~$7.0$~~  & ~~$0.08$~~ & ~~$0.09$~~ & ~~$0.1$~~ \\
$Turb. + error,\ linear\ fit$    & ~~$0.37$~~ & ~~$14.4$~~ & ~~$0.4$~~  & ~~$0.3$~~ & ~~$0.5$~~ \\
$Turb. + error,\ quadratic\ fit$ & ~~$0.47$~~ & ~~$14.4$~~ & ~~$0.4$~~  & ~~$0.3$~~ & ~~$0.6$~~ \\
\\
\hline \hline
\end{tabular}
\caption{Results for the three examples: difference between the reconstructed and the true value of $F_{\theta_1}$, $\Delta F_{\theta_1}$, of the direction of $\vec F$, angle $\epsilon$, and of the position of the source $(\beta_1,\beta_2)$. The corresponding true values of $F_{\theta_1}$ for each example are: 3.3$^{\circ}$ 100 EeV, 0.7$^{\circ}$ 100 EeV and 1.8$^{\circ}$ 100 EeV (1$^{\circ}$ 100 EeV $\approx$ 1.9 $e$ $\mu$G kpc).\label{tabla1}}
\end{table}

In the first example the source is located in galactic coordinates $(l,b)= (220^{\circ},-60^{\circ})$,
corresponding to a region of the sky with a large deflection ($F_{\theta_1}$= 3.3$^{\circ}$ 100 EeV). We see
from Table 1 that the value of $F_{\theta_1}$ is well reconstructed with the linear fit with median relative errors
smaller than 10$\%$. The quadratic fit gives even more accurate results both for the source position and for $F_{\theta_1}$ when no measurement uncertainties are introduced. When the experimental uncertainties (of
the magnitude discussed above) are introduced, the mean reconstructed value of $F_{\theta_1}$, $\left <F_{\theta_1} \right >= (3.0 \pm 0.4)^{\circ}$ 100 EeV, agrees within the error with the true value, $F_{\theta_1}$= 3.3$^\circ$ 100 EeV, in the linear fit case. The quadratic fit leads to a significantly larger uncertainty in the determination of
$F_{\theta_1}$, $\left < F_{\theta_1} \right >= (3.8 \pm 2)^{\circ}$ 100 EeV, that also encompasses the true value of $F_{\theta_1}$. On the other hand, the mean fitted value for the quadratic term, $\left <F_{\theta_1}(\vec{\beta})\frac{\partial F_{\theta_1}}{\partial \theta_1}\big{\arrowvert}_{\vec{\beta}}\right
>= (0.3 \pm 0.4)^\circ$ (100 EeV)$^2$ is compatible with zero. We see that performing a quadratic fit does not lead to an improvement of the reconstruction  in this case for the values of the experimental uncertainties considered, as the addition of an extra parameter just lead to a larger uncertainty in the determination of $F_{\theta_1}$. The position of the source and the direction of $\vec{F}$ are also well reconstructed in all cases. The quadratic fit also leads to an increased uncertainty in $\beta_1$ with respect to the linear fit. In addition, the turbulent component does not have much effect in any result because the regular component gives the dominant contribution to the deflections. In Figure 5, the deflection as a function of $1/E$ is portrayed for the four situations studied in this example.\\

\begin{figure}[!htb]
\subfigure[\label{fig4:a}]{\includegraphics[scale=0.7]{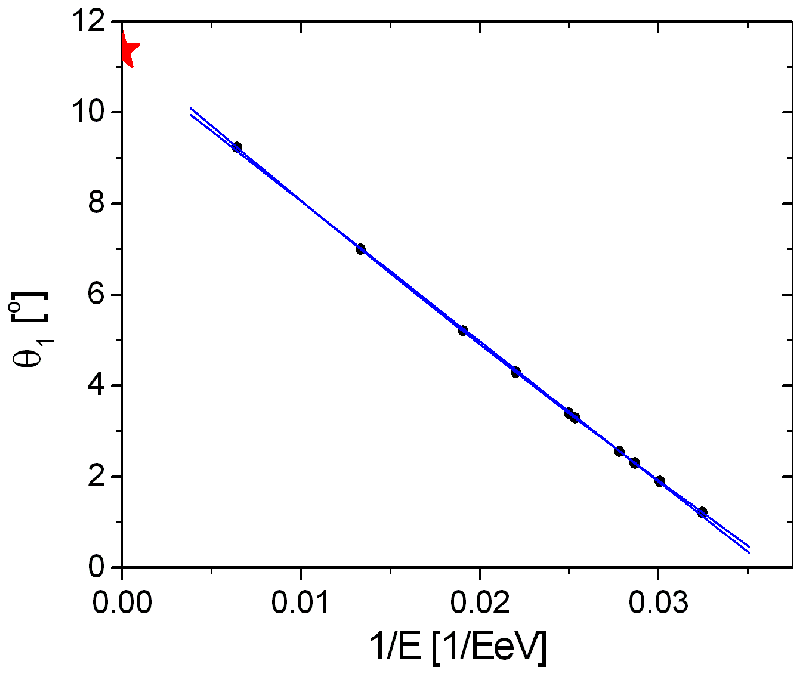}}
\subfigure[\label{fig4:b}]{\includegraphics[scale=0.7]{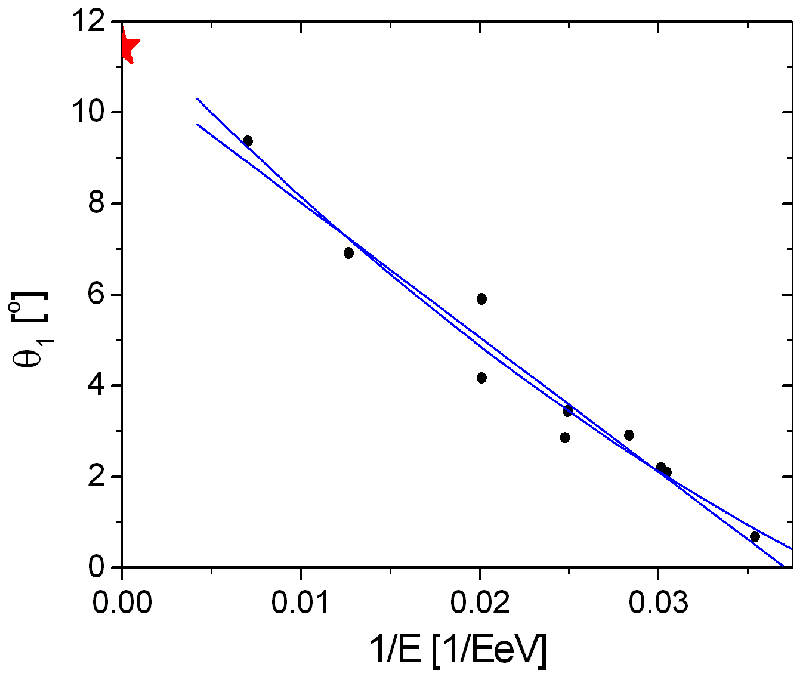}}
\subfigure[\label{fig4:c}]{\includegraphics[scale=0.7]{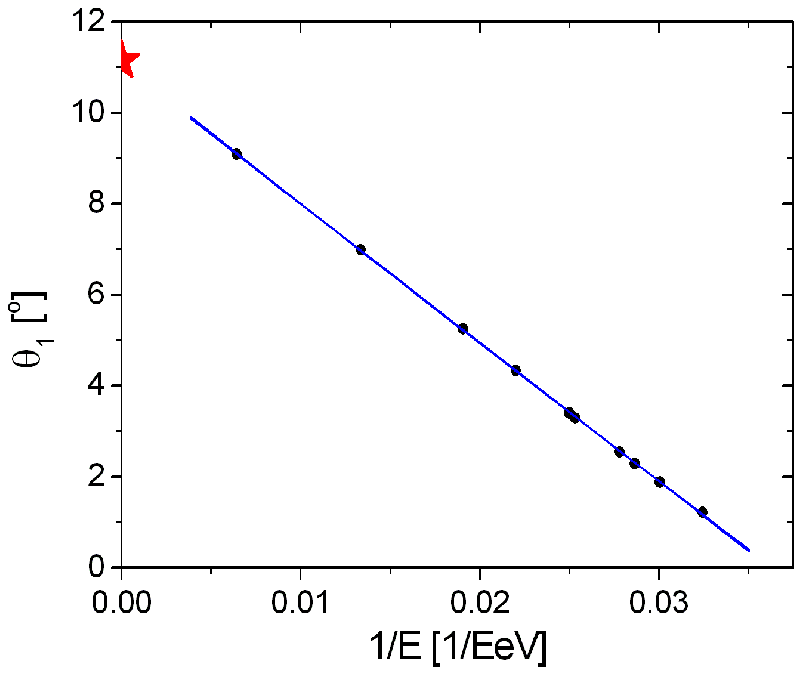}}
\subfigure[\label{fig4:d}]{\includegraphics[scale=0.7]{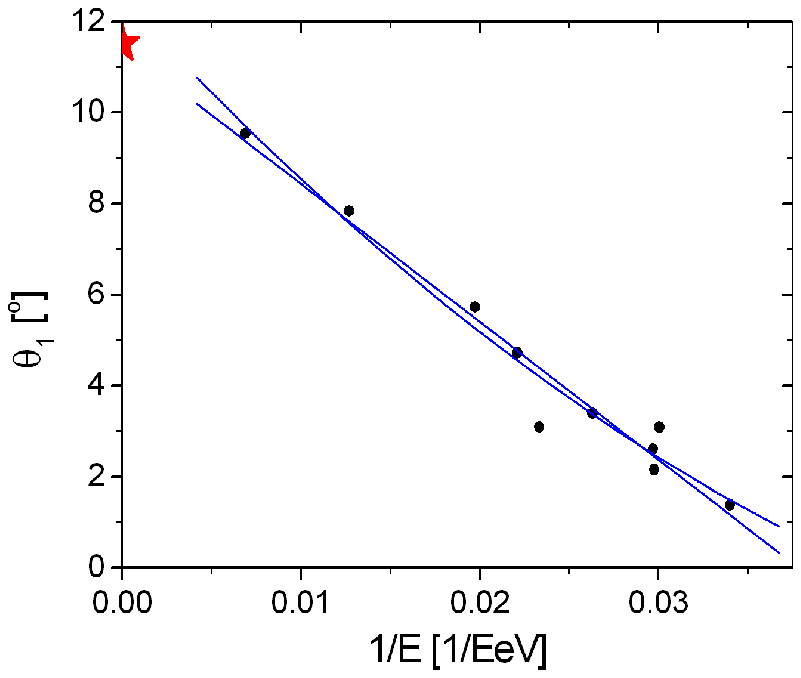}} \caption[]{ Deflection as a function of $1/E$
for a source located at ($l,b$)=($220^{\circ},-60^{\circ}$) considering (a) magnetic field with a
regular component, (b) regular component plus measurement uncertainties in energy and position, (c) regular and
turbulent magnetic field, (d) regular and turbulent component plus measurement uncertainties in energy and
position. The star indicates the true position of the source.}
\end{figure}

For the second example, we consider a source located at $(l,b)= (55^{\circ},20^{\circ})$, corresponding to a
region with a small deflection ($F_{\theta_1}$= 0.7$^{\circ}$ 100 EeV). The error in the value of $F_{\theta_1}$
is comparable to that of the previous example (though the relative error is considerably larger, as
$F_{\theta_1}$ is smaller) and the position of the source is well reconstructed, as it is shown in Table I. Both the mean reconstructed values of $F_{\theta_1}$ ($\left < F_{\theta_1} \right > = (0.9 \pm 0.4)^{\circ}$ 100 EeV) and the source coordinate $\beta_1$ coincide within the error of the fit with the true values. The direction of $\vec F$ has a poorer reconstruction than in the previous example. This happens because as the deflection in this region of the sky is small, there is a smaller lever arm between the highest and the smallest energy events to fix the deflection direction and the uncertainties introduced in the angle are relatively more significant. Although the quadratic fit gives a better reconstruction of $F_{\theta_1}$ when measurement uncertainties are neglected, it is less accurate when adding errors as in the previous example. The turbulent component does not introduce substantial changes either in this case.\\

For the third example we consider a source at $(l,b)= (20^{\circ},-40^{\circ})$, that corresponds to a region
with deflections close to the mean ($F_{\theta_1}$= 1.8$^{\circ}$ 100 EeV) but near to a fold of the sky, as
opposed to the other examples that were located in regions where the sky sheet does not suffer much deformation
when lowering the energy of the incoming particles. For the linear fit, the mean reconstructed value of $F_{\theta_1}$, $\left < F_{\theta_1} \right >= (1.4 \pm 0.4)^{\circ}$ 100 EeV, coincides within the error with the true value. The median difference between the value of $F_{\theta_1}$ found and the real one is 23$\%$ and in the direction of $\vec{F}$ is
16.1$^\circ$. The reconstruction without introducing the experimental errors is less accurate than in the first case because close to the folds the departures from the linear (and quadratic) approximation for the deflection as a function of $1/E$ are larger. The position of the source is reconstructed with a median accuracy of $0.6^\circ$; and the quadratic fit behavior is similar to that in the previous examples.\\

If a random field with  $B_{rms}= 4\ \mu$G instead of $B_{rms}= 2\ \mu$G is considered, multiple images due to
the turbulent component can appear above 30 EeV in the second example (in about half of the different
realizations). However, the accuracy of the reconstruction is not much affected by the turbulent component also
for this larger amplitude of the random field.\\
%%%%%%%%%%%%%%%%%%%%%%%%%%%%%%%%%%%%%%%%%%%%%%%%%%%%%%%%%%%%%%%%%%%%%%%%%%%%%%%%%%%%%%%%%%%%%
\section{Results}
In order to study the uncertainties in the reconstruction of $\vec F$ and in the position of the source we
randomly select the location of 100 sources and perform the analysis described in the previous section to each
one of them. We present the results for only the regular magnetic field component as the effects of the
turbulent field on the accuracy of the reconstruction are small as exemplified in the previous section. The
position of the sources are plotted in Figure 3 together with 10 cosmic rays arrival directions associated to
each one, corresponding to the different cosmic ray energies considered. The results for the accuracy of the
reconstruction are presented in Table II where the median of the difference between the reconstructed and the true value of $F_{\theta_1}$, $\Delta F_{\theta_1}$, of the angle defining the direction of $\vec{F}$, $\epsilon$, and of the position of the source, ($\Delta \beta_1, \Delta \beta_2$), are quoted. The histograms for the full distribution for the 100 source positions and 100 realizations of the measurement uncertainties are plotted in Figures 6, 7 and 8.\\

\begin{table}[!hbtp]
\begin{tabular}{lccccc}
\hline \hline
\\
~~~~       &~~~~$\Delta F_{\theta_1}$~~ & ~~ $\epsilon$~~ & ~~ $\Delta \beta_1$~~ & ~~ $\Delta \beta_2$~~ & ~~ $\mid \Delta \vec{\beta} \mid$~~\\
~~~~       &~~~~$[^{\circ} 100\ {\rm EeV}]$~~ & ~~ $[^{\circ}]$~~ & ~~ $[^{\circ}]$~~ & ~~ $[^{\circ}]$~~ & ~~ $[^{\circ}]$~~\\
\\
\hline
\\
$Regular,\ linear\ fit$          & ~~$0.25$~~ & ~~$3.9$~~ & ~~$0.2$~~  & ~~$0.09$~~ & ~~$0.2$~~\\
$Regular,\ quadratic\ fit$       & ~~$0.05$~~ & ~~$3.9$~~ & ~~$0.02$~~ & ~~$0.09$~~ & ~~$0.09$~~\\
$Reg. + error,\ linear\ fit$     & ~~$0.32$~~ & ~~$5.8$~~ & ~~$0.3$~~  & ~~$0.3$~~ & ~~$0.5$~~\\
$Reg. + error,\ quadratic\ fit$  & ~~$0.65$~~ & ~~$5.8$~~ & ~~$0.5$~~  & ~~$0.3$~~ & ~~$0.6$~~\\
\\
\hline \hline
\end{tabular}
\caption{Results for 100 sources with randomly selected positions. Median of the difference between the reconstructed and the true value of $F_{\theta_1}$, $\Delta F_{\theta_1}$, of the
direction of $\vec F$, angle $\epsilon$, and of the position of the source $\vec \beta=(\beta_1,\beta_2)$,
$\Delta \beta_1$, $\Delta \beta_2$, $\mid \Delta \vec{\beta} \mid$, for the linear and the quadratic fits.\label{tabla2}}
\end{table}

\begin{figure}[!htb]
\subfigure[\label{fig5:a}]{\includegraphics[scale=0.7]{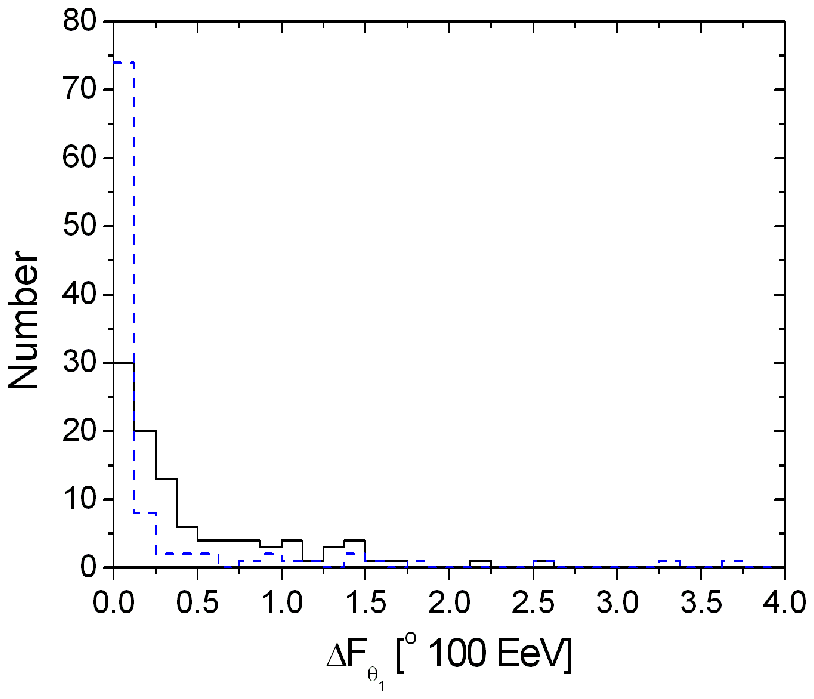}}
\subfigure[\label{fig5:b}]{\includegraphics[scale=0.7]{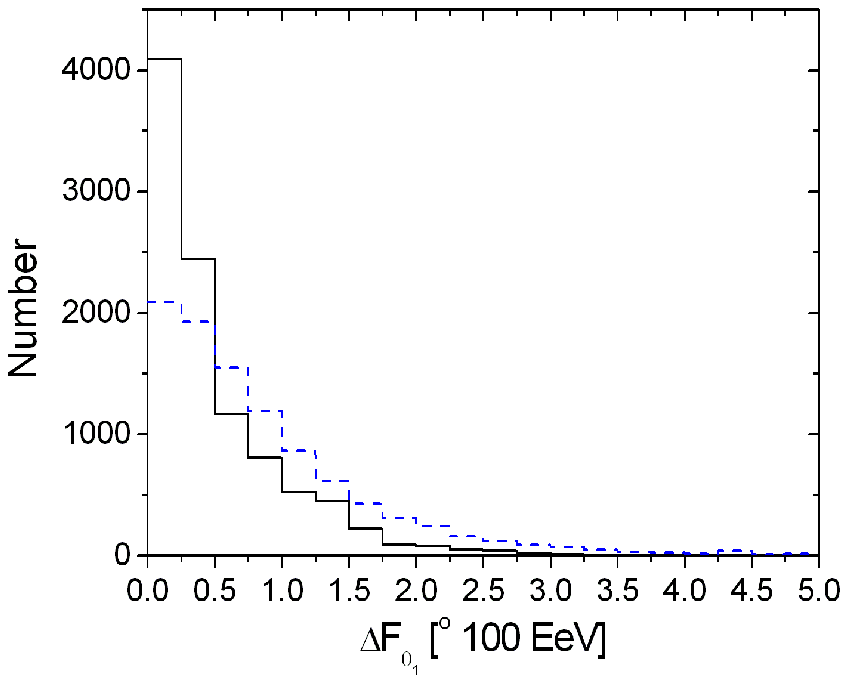}} \caption[]{Histograms of the
difference between the reconstructed and the true value of $F_{\theta_1}$, $\Delta F_{\theta_1}$, for the 100 source positions considering only a regular magnetic field without (a) and
with (b) measurement uncertainties (for the 100 different realizations of the angle and energy uncertainties). Solid lines
correspond to the linear fit and dashed lines to the quadratic fit results.}
\end{figure}

\begin{figure}[!htb]
\subfigure[\label{fig6:a}]{\includegraphics[scale=0.7]{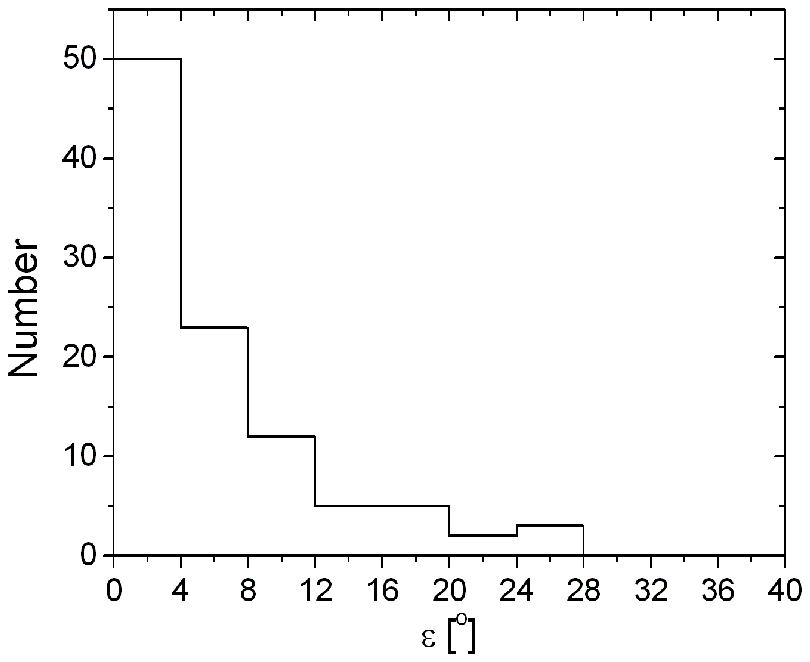}}
\subfigure[\label{fig6:b}]{\includegraphics[scale=0.7]{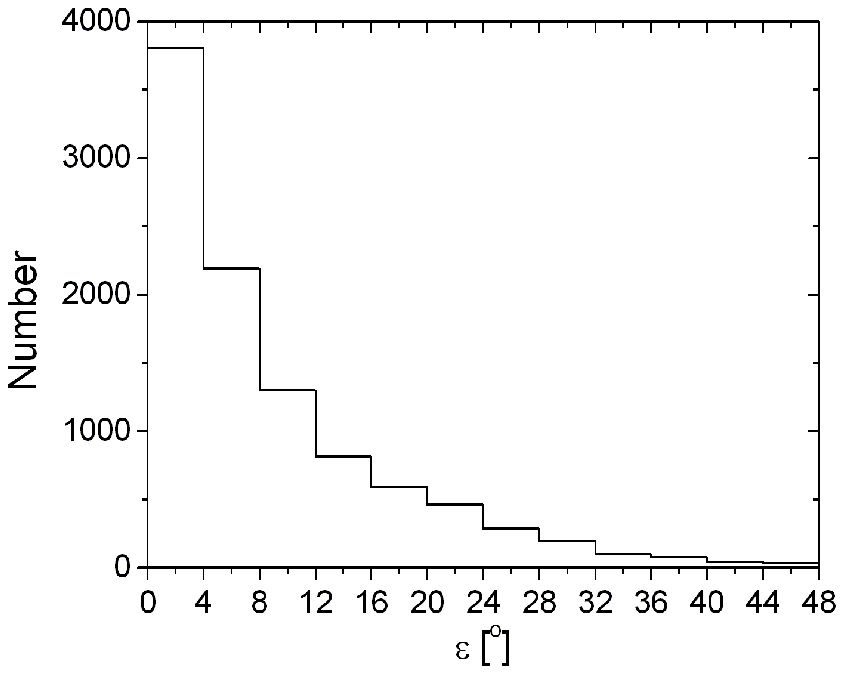}} \caption[]{Histograms of the difference between the reconstructed and the true value of the direction of $\vec F$, $\epsilon$, for the 100 source positions considering only a regular magnetic field without (a) and with (b) measurement uncertainties.}
\end{figure}

Applying a linear fit the reconstruction of $F_{\theta_1}$ is achieved with median difference between the reconstructed and the true value equal to 0.25$^\circ$ 100 EeV and 0.32$^{\circ}$ 100 EeV without and with measurement uncertainties respectively (i.e. 50$\%$ of the realizations have errors smaller than these values). On the other hand, in 57$\%$ of the cases the true value of $F_{\theta_1}$ lies within the error bar of the reconstructed value for the linear fit (in 90$\%$ of the cases it lies within 2.8 times the error bar). The corresponding median errors in the direction of $\vec F$ in these two cases are 3.9$^\circ$ and $5.8^\circ$ respectively.\\

When comparing $\Delta F_{\theta_1}$ obtained with the linear fit and the one with the quadratic fit, the latter
is more accurate when no measurement uncertainties are introduced. However, when experimental errors of the magnitude considered are taken into account, the quadratic fit has larger uncertainties. This behavior is shown clearly in
Figure 6 and in Table II where the median difference between the reconstructed values and the true values of $F_{\theta_1}$ for the quadratic fit grows from 0.05$^{\circ}$ 100 EeV to 0.65$^{\circ}$ 100 EeV when adding the measurement errors. Improving the experimental resolution, the accuracy of the quadratic fit improves, as it is shown in Figure 9. However, only an experiment with very good energy and angular resolution (for example 0.3$^{\circ}$ in angle and 5$\%$ in energy) would have a better accuracy by using the quadratic fit instead of the linear one for a statistics of events comparable to the one considered here.\\

\begin{figure}[!htb]
\subfigure[\label{fig7:a}]{\includegraphics[scale=0.7]{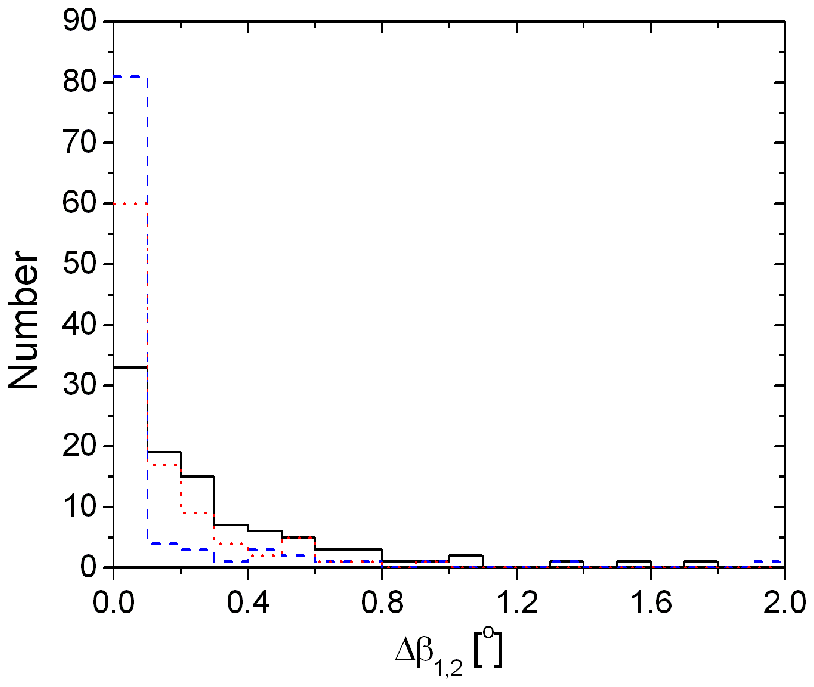}}
\subfigure[\label{fig7:b}]{\includegraphics[scale=0.7]{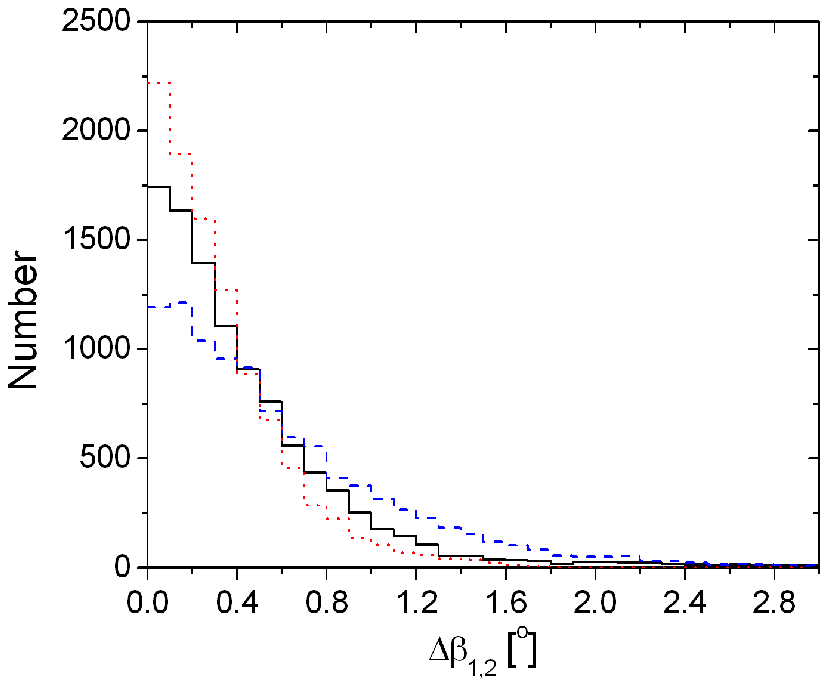}} \caption[]{Histograms of the difference between the reconstructed and the true value of
the position of the source, $(\Delta \beta_1,\Delta \beta_2)$, for the 100 source positions considering only a regular magnetic field without
(a) and with (b) measurement uncertainties. Solid lines correspond to the uncertainty in $\beta_1$ obtained with
the linear fit, dashed lines to $\Delta \beta_1$ obtained with the quadratic fit and dotted lines to the uncertainty in
$\beta_2$.}
\end{figure}

The results for the reconstruction of the source direction are shown in Figure 8. Using the linear fit the
position of the source is obtained with a median error of $0.2^{\circ}$ without introducing the experimental
uncertainties and of $0.5^{\circ}$ when they are included. Once again, the median error applying a quadratic fit
is lower, being $\mid \Delta \vec{\beta} \mid = 0.09^{\circ}$ when no uncertainties are included, but when considering an experiment with an angular resolution of 1$^{\circ}$ and an energy uncertainty of 10$\%$, rises to  $\mid \Delta \vec{\beta} \mid= 0.6^{\circ}$. In 85$\%$ of the cases the true source position is within the error bar of the reconstructed direction in the linear fit case.\\

The fact that the quadratic fit reconstruction accuracy becomes worse than the linear fit one when the
measurement errors are considered can be understood as follows. The mean value of the quadratic term
contribution to the deflection in Eq. (5) for the 100 source directions is $\left
<F_{\theta_1}(\vec{\beta})\frac{\partial F_{\theta_1}}{\partial \theta_1}\big{\arrowvert}_{\vec{\beta}}\right
>= 0.12^\circ$ (100 EeV)$^2$. Thus the expected mean departure from the approximately linear deflection is
smaller than the angular resolution considered ($1^\circ$) up to energies around 35 EeV. Therefore, in most of
the cases the quadratic deflection term is smaller than the angular accuracy and the addition of an extra
parameter to the fit leads to a worse determination of the linear term. Only for energy thresholds significantly
below 35 EeV would the quadratic fit give more accurate results for the magnetic field model considered.\\

Regarding magnetic lensing effects, from the 100 randomly selected sources, 20 of the positions are crossed by a
fold at energies above 30 EeV and this fact affects the reconstruction of $\vec F$ as it is illustrated in the
third example discussed in the previous section. Out of the 20, in 16 cases the fold crosses the position of the
source at energies between 30 and 40 EeV.\\

\begin{figure}[!htb]
\begin{center}
\includegraphics[scale = 0.75]{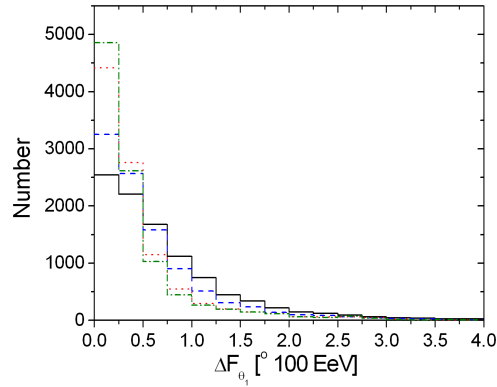}
\caption[]{Histograms of the difference between the reconstructed and the true value of $F_{\theta_1}$, $\Delta F_{\theta_1}$, obtained with a quadratic
fit, considering a regular magnetic field with measurement uncertainties. Solid lines correspond to an angular
resolution of 0.8$^{\circ}$ and an energy uncertainty of 8$\%$, dashed lines to 0.6$^{\circ}$ and 6$\%$, dotted
to 0.4$^{\circ}$ and 4$\%$ and dash-dot lines to 0.3$^{\circ}$ and 5$\%$.}\label{figura 8}
\end{center}
\end{figure}

If one includes a turbulent component of the galactic magnetic field, the results are similar except for some of
the sources near the galactic plane for which the cosmic ray trajectories traverse a larger region with
turbulent magnetic field, leading to multiple images at energies above the 30 EeV threshold considered. This
happens for 14 of the source locations considered. To illustrate this point we plot in Figure 10 the deflection
as a function of $1/E$ for two source positions in the cases in which only the regular component of the magnetic
field is considered and considering both the regular and the turbulent components. The left panel corresponds to
a source located at $(l,b)$=($223.6^{\circ},-50^{\circ}$) while the right panel corresponds to a source near
the galactic plane at $(l,b)$=($117.1^{\circ},1.1^{\circ}$). In the first case, exemplifying sources far from
the galactic plane, the turbulent component only produces small departures of the arrival direction of cosmic
rays  from that resulting from the regular component alone. The source near the galactic plane (right panel)
shows a similar behaviour for energies larger than 75 EeV, but below this energy multiple images appear close to
the principal one. The deflections in the energy range considered are however small and the resulting angular
spread in the event directions is not much larger than the angular resolution considered. If one compares the
accuracy of the linear fit reconstruction for the case of 10 events simulated in the magnetic field with regular
and turbulent components and for the case considering only the regular component, but including the experimental
uncertainties, the results are similar with the exception of the accuracy in the reconstruction of the direction
of $\vec F$, for which the median of $\epsilon$ grows from 5.8$^\circ$ to 12.3$^\circ$. This is due to the fact
that the turbulent field has no preferential direction and hence adds to the integral along the path of the
charged particle for both $F_{\theta_1}$ and $F_{\theta_2}$. However, out of the 14 source positions for which
multiple images above 30 EeV were obtained, the one just considered is the source with the lowest value of $\mid
\vec F \mid$ and therefore corresponds to the worst case.\\

\begin{figure}[!htb]
\subfigure[\label{fig9:a}]{\includegraphics[scale=0.7]{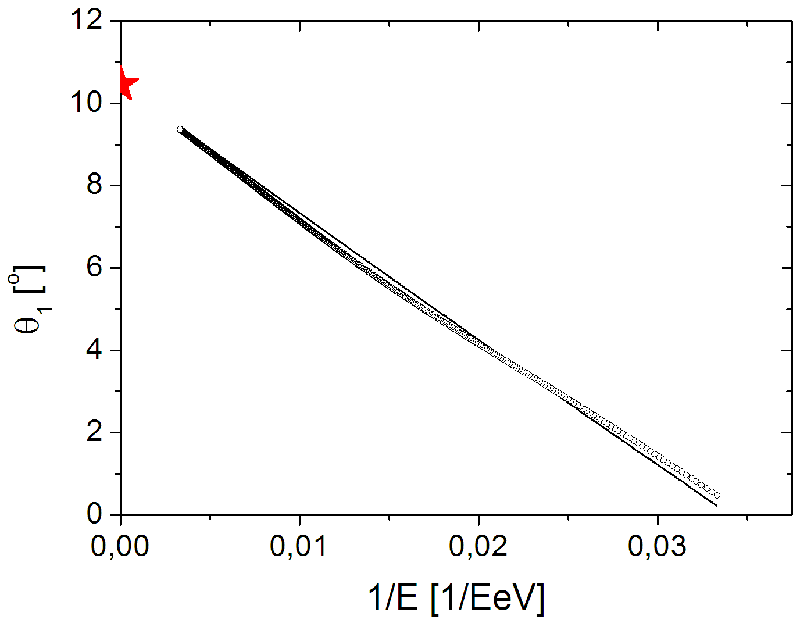}}
\subfigure[\label{fig9:b}]{\includegraphics[scale=0.7]{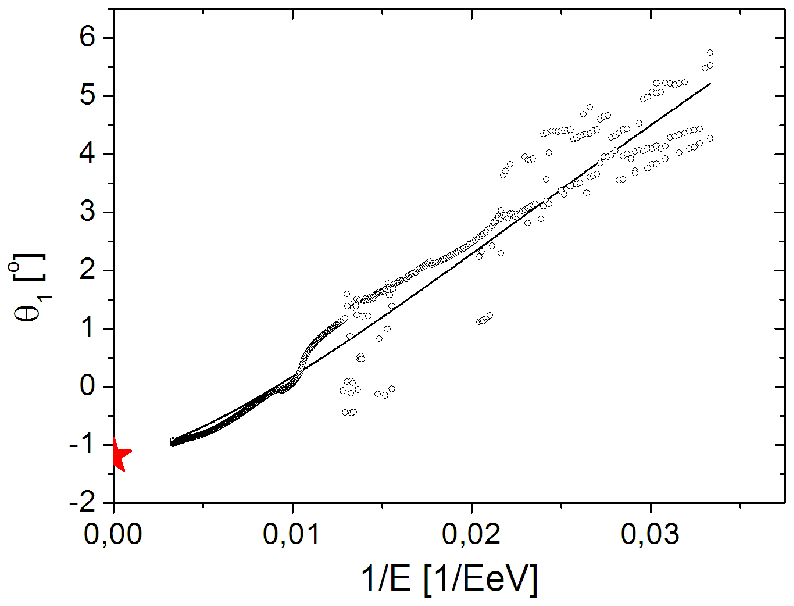}} \caption[]{Deflection as a function of
$1/E$ for two sources: (a) is located at $(l,b)$=($223.6^{\circ},-50^{\circ}$) and (b) is near the galactic
plane at $(l,b)$=($117.1^{\circ},1.1^{\circ}$). The solid line correspond to the deflection caused considering
only a regular component of the galactic magnetic field and empty circles correspond to the deflection caused by both
a regular and a turbulent component. In (b) some of the secondary images of the source due to the turbulent component
are shown. The star indicates the true position of the source.}
\end{figure}

One may wonder if once a set of events from a source is identified it may be more convenient to use only a
subset of the most energetic events for the reconstruction, as the linear relation is more accurate at high
energies. In fact, when no measurement uncertainties are taken into account, the reconstruction of $\vec{F}$ and
the position of the source improves when using only the most energetic events. For instance, if we consider the 5
most energetic events out of the 10 we simulated for each source, corresponding approximately to adopting an
energy threshold of 50 EeV, the linear fit gives in 50$\%$ of the examples accuracies better than: $\Delta F_{\theta_1}$=0.16$^{\circ}$ 100 EeV, $\epsilon$=2.2$^{\circ}$ and $\mid \Delta \vec{\beta} \mid$=0.09$^{\circ}$. Therefore, the reconstruction accuracy is similar as when applying the quadratic fit considering all the events. However, when experimental uncertainties are considered, the median accuracies of the reconstruction applying a linear fit to the 5 most energetic events rise to: $\Delta F_{\theta_1}$=0.46$^{\circ}$ 100 EeV, $\epsilon$=10.5$^{\circ}$ and $\mid \Delta \vec{\beta} \mid$=0.7$^{\circ}$, and we see then that it is best to use all the events.\\

Also one may wonder if it would be preferable to consider a lower threshold for the energy, for example 20 EeV
instead of 30 EeV, allowing to gain more statistics. However, the linear fit reconstruction accuracy is not
improved because the departures from the linear relation between the deflection and $1/E$ are more relevant. For
instance, if we select 16 events above 20 EeV and take into account the experimental resolution, the accuracy
obtained applying the linear fit is in 50$\%$ of the cases better than: $\Delta F_{\theta_1}$=0.41$^{\circ}$ 100
EeV, $\epsilon$=6.0$^{\circ}$ and $\mid \Delta \vec{\beta} \mid$=0.5$^{\circ}$. The accuracy of the
reconstruction is comparable using the linear and quadratic fit for this energy threshold.\\

Another question is how the accuracy of the reconstruction depends on the number of events that are detected
from a source. If instead of 10 events, as considered in the previous analysis, only 5 events above 30 EeV were
detected, the reconstruction accuracy does not change appreciably when measurement uncertainties are not
introduced, but it moderately worsens when uncertainties are taken into account, becoming: $\Delta
F_{\theta_1}$=0.37$^{\circ}$ 100 EeV, $\epsilon$=7.1$^{\circ}$ and $\mid \Delta \vec{\beta}
\mid$=0.6$^{\circ}$.\\

Finally, future satellite experiments such as JEM-EUSO will have larger aperture, gathering hence a larger
number of events, but will also have worse angular and energy resolution than ground based experiments. To study
this difference, we consider an experiment with an angular resolution of 2$^{\circ}$ and an energy uncertainty
of 20$\%$ and multiplets of 20 events above 50 EeV. The median accuracy of the reconstruction for the 100 source
directions considered are in this case: $\Delta F_{\theta_1}$=0.45$^{\circ}$ 100 EeV, $\epsilon$=12.6$^{\circ}$
and $\mid \Delta \vec{\beta} \mid$=0.7$^{\circ}$. Therefore, for the resolution and number of events considered,
the accuracy would be worse than for the case of multiplets of 10 events above 30 EeV with the resolution
associated to ground based experiments.\\

%%%%%%%%%%%%%%%%%%%%%%%%%%%%%%%%%%%%%%%%%%%%%%%%%%%%%%%%%%%%%%%%%%%%%%%%%%%%%%%%%%%%%%%%%%%%%
\section{Conclusions}
We have studied the accuracy with which the position of the source and the integral of the orthogonal component
of the magnetic field can be reconstructed in case that several events from the same source are detected. We have used simulated sets of events arriving to the Earth from randomly located sources in the sky after travelling through the galactic magnetic field that we model with a regular and a turbulent component. Although the amplitude and the direction of the deflections depend on the model considered, the sets of simulated events provide a sample of the realistic types of deflections that can be expected. Reconstruction using a linear and a quadratic fit to the relation between the events position and $1/E$ has been analysed. For the magnetic field model considered and the reference value of 10 detected events with energy above 30 EeV, the median errors of the reconstruction of $F_{\theta_1}$ applying a linear fit are $0.25^{\circ}$ 100 EeV when no experimental uncertainty is introduced and $0.32^{\circ}$ 100 EeV when a $1^\circ$ uncertainty in the position and a $10\%$ uncertainty in the energy are considered (1$^{\circ}$ 100 EeV $\approx$ 1.9 $e$ $\mu$G kpc). Furthermore, the direction of $\vec F$ is obtained with a median error of $3.9^{\circ}$ and $5.8^{\circ}$ without and with experimental resolution respectively, while the position of the source is obtained with a median error of $0.2^{\circ}$ and $0.5^{\circ}$ without and with experimental uncertainty and applying a linear fit.\\

We found that the quadratic fit gives more accurate results than the linear fit when no measurement errors are
introduced. However, for the magnitude of the experimental uncertainties considered, the linear fit is more
accurate than the quadratic one when the experimental errors are taken into account. Only an experiment with
very good energy and angular resolution (for example 0.3$^{\circ}$ in angle and 5$\%$ in energy) would have a
better accuracy by using the quadratic fit instead of the linear one with 10 events from the same source. If
more than 10 events above 30 EeV from a source were detected, the reconstruction accuracy is improved in a
greater relative measure for the quadratic fit than for the linear one. For example, for 25 or more detected
events, an angular resolution of $0.5^\circ$ and an energy resolution better than $10\%$, the quadratic fit
results become more accurate than the linear ones. In general, we can say that the quadratic fit is preferable whenever the value of the quadratic term is incompatible with zero within the error bar of the fit.\\

The turbulent component of the galactic magnetic field does not have a significant effect in the reconstruction
accuracy, except for some sources near the galactic plane that have multiple images at higher energies than when
considering only the regular component. At these energies the secondary images appear near the principal one
when comparing to the experimental uncertainties considered here and the effect on the reconstruction accuracy
is not large.\\

These results show that once several events from the same source are detected, it will be possible to
reconstruct the source position with a good accuracy. It will also be possible to measure the integral of the
orthogonal component of the magnetic field in the direction of the source. This will nicely complement the
rotation measure observations that provide the integral of the parallel component of the magnetic field, giving
hence a new insight into this open astrophysical problem.\\

\section{Acknowledgments}
This work is supported by ANPCyT (grant PICT 13562-03) and CONICET (grant PIP 5231). We thank P. L. Biermann and B. Baughman for comments.\\

%%%%%%%%%%%%%%%%%%%%%%%%%%%%%%%%%%%%%%%%%%%%%%%%%%%%%%%%%%%%%%%%%%%%%%%%%%%%%%%%%%%%%%%%%%%%%%

%%%%%%%%%%%%%%%%%%%%%%%%%%%%%%%%%%%%%%%%%%%%%%%%%%%%%%%%%%%%%%%%%%%%%%%%%%%%%%%%%%
%%%%%%%%%%%%%%%%%%%%%%%%%%%%%%%%%%%%%%%%%%%%%%%%%%%%%%%%%%%%%%%%%%%%%%%%%%%%%%%%%%%%%

\begin{thebibliography}{99}
\bibitem {science} J. Abraham et al [The Pierre Auger
Collaboration], \emph{ Correlation of the highest energy cosmic rays with nearby extragalactic objects},
\emph{Science} \textbf{318} (2007) 938-943 [astro-ph/0711.2256].
\bibitem{uchihori} Y. Uchihori, M. Nagano, M. Takeda, M. Teshima, J. Lloyd-Evans, A.A. Watson, \emph{Cluster analysis of extremely high-energy cosmic rays in the northern sky}, \emph{Astropart. Phys.} \textbf{13} (2000) 151 [astro-ph/9908193].
\bibitem{stanev} T. Stanev,  \emph{Ultra high energy cosmic rays and the large scale structure of the galactic magnetic field}, \emph{Astrophys. J.} \textbf{479} (1997) 290 [astro-ph/9607086].
\bibitem{tanco} G. Medina Tanco, E. De Gouveia dal Pino, J. Horvath, \emph{Deflection of ultra-high energy cosmic rays by the galactic magnetic field: from the sources to the detector}, \emph{Astrophys. J.} \textbf{479} (1998) L200 [astro-ph/9707041].
\bibitem {toes} D. Harari, S. Mollerach, E. Roulet, \emph{The toes of the
ultra high energy cosmic ray spectrum}, \emph{J. High Energy Phys.} \textbf{08} (1999) 022 [astro-ph/9906309].
\bibitem{tinyakov} P.G. Tinyakov, I.I. Tkachev, \emph{Tracing protons through the galactic magnetic field: A Clue for charge composition of ultrahigh-energy cosmic rays}, \emph{Astropart. Phys.} \textbf{18} (2002) 165 [astro-ph/0111305].
\bibitem{prouza} M. Prouza, R. Smida, \emph{The galactic magnetic field and propagation of ultrahigh energy cosmic rays}, \emph{Astron. Astrophys}. \textbf{410} (2003) 1 [astro-ph/0307165].
\bibitem{takami} H. Takami, K. Sato, \emph{Distortion of ultrahigh-energy sky by Galactic Magnetic Field},
\emph{Astrophys. J.} \textbf{681} (2008) 1279 [astro-ph/0711.2386].
\bibitem{alvarez}  J. Alvarez-Mu\~niz, R. Engel, T. Stanev, \emph{Ultrahigh-energy cosmic ray propagation in the galaxy: Clustering versus isotropy}, \emph{Astrophys. J.} \textbf{572} (2001) 185 [astro-ph/0112227].
\bibitem {sign} D. Harari, S. Mollerach, E. Roulet, \emph{Signatures of galactic magnetic lensing upon
ultra high energy cosmic rays}, \emph{J. High Energy Phys.} \textbf{02} (2000) 035 [astro-ph/0001084].
\bibitem {turb} D. Harari, S. Mollerach, E. Roulet and F. Sanchez, \emph{Lensing of ultra high
energy cosmic rays in turbulent magnetic fields}, \emph{J. High Energy Phys.} \textbf{03} (2002) 045
[astro-ph/0202362].
\bibitem {spec} D. Harari, S. Mollerach, E. Roulet, \emph{Astrophysical magnetic field reconstruction and
spectroscopy with ultra high energy cosmic rays}, \emph{J. High Energy Phys.} \textbf{07} (2002) 006
[astro-ph/0205484].
\bibitem {han1} J. L. Han, \emph{Magnetic structure of our Galaxy: A review of observations}, IAU Symp.259 `Cosmic Magnetic Fields: From Planets, to Stars and Galaxies' Proceedings (2009) [astro-ph/0901.1165].
\bibitem{beck} R. Beck, \emph{Galactic and extragalactic magnetic fields},
astro-ph/0810.2923.
\bibitem {han2} H. Men, K. Ferriere and J. L. Han, \emph{Observational constraints on models for the interstellar magnetic field in the Galactic disk}, (2008) [astro-ph/0805.3454].
\bibitem{heiles} C.Heiles, \emph{The Local Direction and Curvature of the Galactic Magnetic Field Derived from Starlight Polarization}, \emph{Astrophys. J.} \textbf{462} (1996) 316.
\bibitem{han3} J. L. Han, R. N. Manchester, A. G. Lyne, G. J. Qiao and W. van Straten, \emph{Pulsar rotation measures and the large-scale structure of Galactic magnetic field}, \emph{Astrophys. J.} \textbf{642} (2006) 868-881 [astro-ph/0601357].
\bibitem{beck2} R. Beck, A. Shukurov, D. Sokoloff, R. Wielebinski, \emph{Systematic bias in interstellar magnetic field estimates}, \emph{Astron. Astrophys.} \textbf{411} (2003) 99-107 [astro-ph/0307330]
\bibitem{cowsik} R. Cowsik, J. Mitteldorf, \emph{Turbulence-enhanced synchrotron radiation in the Galaxy}, \emph{Astrophys. J.} \textbf{189} (1974) 51.
\bibitem{sarkar} S. Sarkar, \emph{Does the galactic synchrotron radio background originate in old supernova remnants?}, \emph{Mon. Not. R. Astron. Soc.} \textbf{199} (1982) 97.
\bibitem {rand} R. J. Rand, S.R. Kulkarni, \emph{The Local Galactic Magnetic Field}, \emph{Astrophys. J.}
\textbf{343} (1989) 760-772.
\bibitem {ohno} H. Ohno, S. Shibata, \emph{The Random Magnetic Field in the Galaxy}, \emph{Mon. Not.
R. Astron. Soc.} \textbf{262} (1993) 953-962.
\bibitem {mst} D. Harari, S. Mollerach, E. Roulet, \emph{Detecting filaments in the ultra high energy cosmic ray
distribution}, \emph{Astropart. Phys.} \textbf{25} (2006) 412 [astro-ph/0602153].
\end{thebibliography}
\end{document}